\documentstyle[aps,prb,preprint,floats]{revtex}

\begin{document}

\draft
\title{Thermodynamics and Kinetics of Folding of a Small Peptide}   
\author{Ulrich H.E. Hansmann \footnote{hansmann@mtu.edu}}
\address{Department of Physics, Michigan Technological University,
         Houghton, MI 49931-1291, USA}
\author{Jose N. Onuchic \footnote{jonuchic@ucsd.edu}}
\address{Department of Physics, University of California at San Diego,
         La Jolla, CA 92093-0319,USA}
\date{\today}
\maketitle
\begin{abstract}
We study the thermodynamics and kinetics of folding for a small peptide. 
Our data rely on Monte Carlo simulations where the interactions among 
all atoms are taken into account. Monte Carlo kinetics is used to study
folding of the peptide at suitable temperatures. The results of these canonical
simulations are compared with that of  a generalized-ensemble 
simulation.  Our work demonstrates that 
concepts of folding which were developed in the past for minimalist
models hold also for this peptide when simulated with an all-atom
force field.
\end{abstract}

The importance of understanding the statistical physics of the
protein-folding problem has been  recently 
recognized.\cite{ONGCS,DC,SHA4,BOSW,Gruebele,SB,GTh} 
The most prominent example is the  
 energy landscape theory and the funnel 
concept.\cite{Bryngelson87,LMO,Onuchic97}
Originaly developed out of 
analytical and numerical investigations of minimal protein
models (which capture only a few, but supposedly dominant parameters 
 in real proteins), 
such concepts were subsequently also used to probe 
the folding properties of more realistic models of proteins 
where the interactions between all
atoms were taken into account.\cite{Brooks,Brooks2,HMO97b,HOO98b}
One example is our previous work on the  pentapeptide Met-enkephalin 
for which we found indeed a funnel-like structure of the 
free energy landscape and were able to determine
its characteristic temperatures of folding.\cite{HMO97b,HOO98b}
In this paper we  complement the 
the above research on the thermodynamic of Met-enkephalin by 
investigating in addition  its kinetics of foldings.

The linear peptide Met-enkephalin has the amino-acid sequence 
Tyr-Gly-Gly-Phe-Met, and at low temperatures  one finds for this
peptide two major groups of well-defined compact structures
which are characterized (and stabilized) by specific hydrogen bonding
 patterns. Structure A is the ground-state conformation in ECEEP/2 and  
has a Type II' $\beta$-turn
between the second and last residue, stabilized by two possible
hydrogen bonds.  The structure B, the second-lowest energy state, is
characterized by  hydrogen bond between Tyr-1 and Phe-4
resulting in a Type-II $\beta$-turn between the first and
fourth residue. The overlap of a given configuration
with the ground state (structure A) and the second-lowest-energy
state (structure B), respectively, allows to distinguish between
the various  low-energy conformations and defines in a natural way 
two order parameters for our system.

Simulation of realistic protein models where the interaction between all
atoms are taken into account are extremely difficult. Because of the
rough energy landscape, simulations based on canonical Monte Carlo or
molecular dynamics techniques will get trapped at low temperatures in
one of the multitude of local minima separated by high energy
barriers. One possibility to overcome these difficulties are 
{\it generalized-ensemble} techniques.\cite{HO96b,OurReview98}  
Their usefulness for calculation of thermodynamic averages was 
extensively tested and demonstrated in protein 
simulations.\cite{HO,OH95b,HO98d}  For a recent review see, for 
instance, Ref.~\onlinecite{OurReview98}. We have used one of these techniques,
described first in  Ref.~\onlinecite{HO96d}, to  study the thermodynamics of
folding of Met-enkephalin,\cite{HOO98b} and utilize it here again 
to calculate the free energy landscape of the peptide as a function 
of our two order parameters.  However, it  is a characteristic of all 
generalized ensemble algorithms  that they  
do not allow to study the kinetic aspects of folding. This is because the 
dynamics of these algorithms are artificial ones. 
While constructed in such a way that the 
correct distribution of states at a certain temperature can be recovered
(and therefore thermodynamic averages calculated), the time evolution
of the system will be different from the one observed in a  
molecular dynamic simulation at that specific temperature.
 For this reason, generalized-ensemble algorithms are
not suitable for a direct investigation of the kinetics of
folding  and we  had to fall back on canonical Monte Carlo simulations.
However, a careful analysis of the free energy landscape of the peptide,
as obtained by our generalized-ensemble method, allows to extract
indirect information on the kinetic of folding which can then be compared
with the outcome of the canonical simulations. Actually, only by using
the results of of the generalized-ensemble simulations we were
able to find the relevant temperatures on which such research of the 
kinetics should focus and  where at the same time 
canonical investigations are still feasible. In that sense, 
our two simulation techniques are complementary.

Our investigation of Met-enkephalin is based on a detailed, all-atom 
representation of that peptide. The interaction between the atoms is
described by a standard force field, ECEPP/2,\cite{EC}  (as implemented
in the KONF90 program\cite{Konf}) and is given by:
\begin{eqnarray}
E_{tot} & = & E_{C} + E_{LJ} + E_{HB} + E_{tor},\\
E_{C}  & = & \sum_{(i,j)} \frac{332q_i q_j}{\epsilon r_{ij}},\\
E_{LJ} & = & \sum_{(i,j)} \left( \frac{A_{ij}}{r^{12}_{ij}}
                                - \frac{B_{ij}}{r^6_{ij}} \right),\\
E_{HB}  & = & \sum_{(i,j)} \left( \frac{C_{ij}}{r^{12}_{ij}}
                                - \frac{D_{ij}}{r^{10}_{ij}} \right),\\
E_{tor}& = & \sum_l U_l \left( 1 \pm \cos (n_l \chi_l ) \right).
\end{eqnarray}
Here, $r_{ij}$ (in \AA) is the distance between the atoms $i$ and $j$, and
$\chi_l$ is the $l$-th torsion angle.
We further fix the peptide bond angles $\omega$
 to their common value $180^{\circ}$, which
 leaves us with 19 torsion angles ($\phi,~\psi$, and $\chi$) as independent
degrees of freedom (i.e., $n_F = 19$).
In our simulations we did not  explicitly include the interaction
of the peptide with the solvent and set the dielectric constant $\epsilon$
equal to 2.

Our simulations were started from  completely random initial
conformations (Hot Start) and  one Monte Carlo sweep updates every torsion 
angle of the peptide once. In our generalized-ensemble algorithm,
configurations are updated according to the following probability weight:
\begin{equation}
w(E) = \left(1+ \frac{\beta (E-E_0)}{n_F}\right)^{-n_F}~,
\label{eqwe}
\end{equation}
where $E_0$ is an estimator for the ground-state energy, $n_F$ is
the number of degrees of freedom of the system, and $\beta = 1/k_BT$
is  the inverse temperature   ($k_B$ is the
Boltzmann constant and $T$ the temperature of the system).
A simulation with such a weight samples a large range of energies, since
high energies are suppressed only by a power law. Hence, the
thermodynamic average of any physical quantity $\cal{A}$
can be calculated over a wide temperature range by \cite{FS}
\begin{equation}
<{\cal{A}}>_T ~=~ \frac{\displaystyle{\int dx~{\cal{A}}(x)~w^{-1}(E(x))~
                 e^{-\beta E(x)}}}
              {\displaystyle{\int dx~w^{-1}(E(x))~e^{-\beta E(x)}}}~,
\label{eqrw}
\end{equation}
where $x$ stands for configurations. It is known from our previous work 
that the ground-state conformation for Met-enkephalin has the
KONF90 energy value
$E_{GS} = -12.2$ kcal/mol.\cite{HO94_3}  We therefore set $E_0 = -12.2$
kcal/mol, $T = 50$ K (or, $\beta = 10.1$ $[\frac{1}{{\rm kcal}/{\rm mol}}]$)
and $n_F =19$ in our probability weight factor in Eq.~(\ref{eqwe}).
All thermodynamic quantities were then  calculated from
 a single production run of 1,000,000 MC sweeps which followed 10,000
sweeps for thermalization. At the end of every fourth sweep we stored
the energies of the conformation, the corresponding (solvent excluded)
volume (in \AA$^3$)
which is calculated by the  double cubic lattice method,\cite{ELASS} 
 and our two ``order parameters''
(the overlap $O_A$ of the conformation with
the (known) ground state (structure A) and the overlap $O_B$ of the
 conformation with conformer B).
The overlap was defined by
\begin{equation}
O(t) = 1 -\frac{1}{90~n_F} \sum_{i=1}^{n_F} |\alpha_i^{(t)}- \alpha_i^{(RS)}|~,
\label{eqol}
\end{equation}
where $\alpha_i^{(t)}$ and $\alpha_i^{(RS)}$ (in degrees) stand for
the $n_F$ dihedral angles of the conformation at $t$-th Monte Carlo sweep
and the  reference state conformation, respectively. Symmetries
for the side-chain angles were taken into account and the difference
$\alpha_i^{(t)}- \alpha_i^{(RS)}$ was always projected into the interval
$[-180^{\circ},180^{\circ}]$. Our definition  guarantees that we have
\begin{equation}
0 \le ~<O>_T~ \le 1~.
\end{equation}

Using the results of our generalized-ensemble simulation, we explored for
various temperatures the free energies
\begin{equation}
G(O_A,O_B) = -k_B T \log P(O_A,O_B)~.
\end{equation}
Here, $P(O_A,O_B)$ is the probability to find a peptide conformation 
with values $O_A$, $O_B$ (at temperature $T$). We chose the normalization 
so that the lowest value
of $G(O_A,O_B)$ is set to zero for each temperature.

The generalized-ensemble simulation was complemented by  100 canonical
Monte Carlo simulations for each chosen temperature. 
In these canonical simulations we measured   
the overlap $O_A$ with the (known) ground  state (Conformer A) after 
each MC sweep. Once we found a configuration with $O_A(t)\ge 0.8$, 
we identified this configuration with the ground state, 
stored the number of MC sweeps $t^{last}$ for further analysis, 
and stopped the MC run.  We also stopped the MC run if no ground state
was sampled after 200,000 MC sweeps, 
i.e. $t^{last}=200,000$ indicates that the ground state was not found in that 
specific run. The frequency $n_f$ with which the ground state 
(conformer A) was sampled in all 100 canonical runs 
and the average minimal folding time $t_f$
defined by
\begin{equation}
 t_f = \frac{1}{100} \sum_{i=1}^{100} t^{last}_i~.
\end{equation}
are other quantities which  we recorded for each temperature. 
Note that $t_f$ is only a lower bound for the folding time since we
stopped our runs after 200,000 sweeps if no ground state was found in a
MC run. For each temperature, we also measured 
the frequency of unfolded states $P_{uf} (t)$
as a function of time (in Monte Carlo sweeps). This quantity is
defined by
\begin{equation}
P_{uf} (t) = \frac{1}{100} \sum_{i=1}^{100} \theta_i (t,t_i^{last})
\label{Puf}
\end{equation}
with
\begin{equation}
\theta_i (t,t_i^{last}) = \left\{ \begin{array}{ll}
                                  1 & t \le t_i^{last}\\
                                  0 & {\rm otherwise}
                                  \end{array}
                          \right.
\end{equation}
In addition to the above quantities we finally  measured  the 
frequency $n_f^{noB}$  with which
the simulations went straight into the ground state (conformer A)
(i.e. without being trapped first in conformer B);  the frequency $n_f^{B}$
of runs which were first trapped in  conformer B before finally finding 
the ground state ($n_f^B$); and the corresponding folding times   
 $t_f^{noB}$  and $t_f^B$. All 6 quantities are listed in Table~1.

For each temperature, we allowed  at least 10 of the 100 canonical runs 
to finish the whole 200,000 sweeps independently on
whether the ground state was found or not. For these 10 runs we 
stored the ``time series'' of conformations for a more detailed analysis. 
This allowed us to calculate the escape (or ``life'') time $\tau_{es}$ 
for conformer A and conformer B. For this we  recorded 
the frequency $n_{es}^A(t)$ ($n_{es}^B(t)$) 
with which the simulations remained in state A (B) for $t$ MC sweeps.
For three out of our four 
temperatures  these frequencies could be fitted by a single exponential fit
\begin{equation}
 n_{es} (t) = A \times e^{-t/\tau_{es}}
\end{equation}
and the resulting escape times $\tau_{es}$  are listed in Table~2.

We start our analyses with the results of our generalized-ensemble simulation.
In earlier work we could  determine  the
collapse temperature $T_{\theta}=295\pm 30$ K and the folding temperature
$T_f = 230 \pm 30$ K.\cite{HMO97b}   While these two characteristic
temperature are important to understand the mechanism of folding of 
Met-enkephalin, 
the essence of the funnel landscape idea is competition between the
tendency towards the folded state and trapping due to ruggedness of
the landscape. One way to measure this  competition is by the
ratio:\cite{tang} 
\begin{equation}
Z = \frac{\overline{E-E_0}}{\sqrt{\overline{E^2} - \bar{E}^2}}~,
\label{eq_Q}
\end{equation}
where the bar denotes averaging over compact configurations.
The landscape theory asserts that good folding protein sequences
are characterized by  large values of $Z$.\cite{tang}  Using the results
of our previous simulations and defining a compact structure as one where
$V(i) \le  1380 \AA^3$, we find $\overline{E-E_0} = 40.7(1.0)$ Kcal/mol,
$\overline{E^2} - \bar{E}^2 = 15.7(2.0)$ (Kcal/mol)$^2$, 
from which we estimate for
 the above ratio $Z=10.3(1.1)$. This value indicates that Met-enkephalin is 
a good folder.

Another way of characterizing the competition between tendency towards
the folded state and roughness of the energy landscape relies on knowledge
of the glass temperature $T_g$. It is expected
that for a good folder the glass transition
temperature, $T_g$, where glass behavior
sets in, has to be significantly lower than the folding temperature
$T_f$, i.e. a good folder can be characterized by the
relation \cite{BOSW}
\begin{equation}
\frac{T_f}{T_g} > 1~.
\end{equation}
We can calculate a crude estimate of the glass transition temperature 
by using the approximation \cite{BOSW}
\begin{equation}
T_g = \sqrt{\displaystyle{\frac{\overline{E^2} -\bar{E}^2}{2 k_B^2 S_0}}}~,
\end{equation}
where the bar indicates again averaging over compact structures
and $S_0$ is the entropy of these states estimated by the relation
\begin{equation}
 S_0 = \log \overline{n}_{compact}  -C
\end{equation}
Here,  $C$ is chosen such that
the entropy of the ground state becomes zero. The results of our simulation 
leads to a value of $s_0 = 55(4)$. Together with the above quoted value
for $\overline{E^2} -\bar{E}^2 = 15.7(2.0)$ (in (Kcal/mol)$^2$  one
 finds as an estimate for the glass transition temperature
\begin{equation}
 T_g = 190 (20)~{\rm K}~.
\end{equation}
This result is in good agreement with a recent estimate $T_g = 180 \pm 30$ K
for that peptide  determined from the change in 
the fractal dimension of the free energy landscape with
temperature.\cite{AH99d} 
Since it was stated in earlier work \cite{HMO97b} that $T_f=230(30)$ K,
it is obvious that the ratio $T_f/T_g > 1$, and again one
finds that Met-enkephalin has good folding properties. We remark
that our results are consistent with an alternative characterization of
folding properties by Thirumalai and collaborators \cite{CTh,KTh} as
was pointed out in detail in Ref.~\onlinecite{HMO97b}.

An advantage of the generalized-ensemble approach is that it allows us to
observe directly the  folding funnel of Met-enkephalin. Our results are 
compared with that of the canonical runs.
In Fig.~1 we show the free energy landscape as a function of 
both the overlap $O_A$ with the ground state and the
overlap $O_B$ with structure B in  the high-temperature
situation (at $T=1000$ K). The free energy has its minimum at small values
of the overlap indicating that both conformers appear with only very
small frequency at high temperature. We have superimposed on the free-energy
landscape,  as calculated from the generalized-ensemble simulation, the
 folding trajectory of a canonical Monte Carlo simulation (marked by dots)
at the same  temperature. However, we did not connect the dots, for otherwise
the plot would become unreadable. It is obvious that the
concentration of the dots marks the time
the simulation spent in a certain region of the landscape. We see that this
time is strongly correlated with the free-energy as calculated from the
generalized-ensemble simulation. For instance, we have no dots for
$O_A \approx 1$ (i.e. the ground state region), a region of the energy
landscape suppressed by many $k_BT$. Actually, the folded state was
found even at this high temperature, in 47 of the 100 canonical runs   
within the available 200,00 sweeps. However,  the ground state is 
not stable at this high temperature. We found that the average
escape time out of this state was  about $\tau_{es}=9$ MC sweeps, which one 
has to compare
with an average minimal folding time of $t_f=145,599$ MC sweeps. 
Hence, the probability to find folded states at such a high temperature 
in canonical simulations is negligible. This results  
is consistent with the displayed free energy landscape
calculated from generalized-ensemble simulation.

At $T=300 K$, which is essentially
the collapse temperature  $T_{\theta} = 295\pm 30$ K  of 
Ref.~\onlinecite{HMO97b},  a large part of the space of possible
configurations  lies
within the $2 k_BT$ contour as is clear from Fig.~2. Correspondingly, 
the dots, which mark the folding trajectory of a canonical simulation at this
temperature, are equally distributed over the whole plot. We remark that
at this temperature the folded conformation was found in all of the 100 
canonical simulations. Compared with the high temperature
$T=1000$ K  the average folding time decreased  by an order of magnitude 
to $t_f=19864$  MC sweeps and  the escape time increased by an 
order of magnitude
towards $\tau_{es}=2000$ MC sweeps enhancing the probability to 
find that state at
$T=300$ K. 

At the folding
temperature $T_f = 230$ K a funnel in the energy landscape appears
with a gradient towards the ground state, but Fig.~3 shows that there
are various other structures, the most notable of which is Conformer B
(where $O_B \approx 1$), with free energies 3 $k_BT$ higher than the
ground-state conformation but separated from each other and the ground
state only by free energy barriers less than 1 $k_BT$.  No other
long-lived traps are populated. Hence, the funnel at $T_f$ is
reasonably smooth.  Folding routes  include direct conversion
from random-coil conformations into Conformer A or some short
trapping in Conformer B region before reaching Conformer A region,
but at the folding
temperature it is possible to reach the ground state from any
configuration without getting  kinetically trapped. This was indeed
observed by us  in the 100 canonical runs we performed at this temperature. 
Some of the runs went directly 
from the unfolded state to the  folded conformation (state A),
while in other runs we saw first short trapping in the region of conformer B
before folding into the  ground-state structure. The folding trajectory
tracjectory  displayed in the figure is an example for  the later case. We 
found as escape time out of conformer B roughly $\tau^B_{es}=5000$ MC sweeps. 
Due to such trapping
only  86 of 100 MC runs found the folded state within 200,000 sweeps,
leading to an average minimal folding time of $t_f=77,230$ MC sweeps,
which is 4 times as long as for the collapse temperature $T_{\theta}=300$ K. 
However,
the escape time for the folded state also increased  to $\tau_{es}^A=19430$
 MC sweeps, about 10 times as long as for $T=300$ K. Hence, 
the interplay of folding time and  life time of the folded state
leads to the increases probability of the folded state in the free
energy plot for this temperature.  

Finally,
Fig.~4 shows the situation for $T= 150$ K where we expect onset of glassy
behavior.  Again one sees a  funnel-like bias toward the ground
state, however, the funnel is no longer smooth and the free energy
landscape is rugged. Free energy barriers of many $k_BT$ now separate
different regions and would act as  long-lived kinetic traps in a
canonical simulation rendering folding at this temperature extremely
difficult. This can be seen for the folding trajectory we display in that
figure: the simulation got trapped in a region of the landscape far away from
the folded state and never reached the folded state within the 200,000 sweeps
of the simulation. Actually, only in 19 out of 100 Monte Carlo simulations of
200,000 sweeps we found the folded state and the minimal folding time is 
at least $t_f=172866$ MC sweeps. 
Our data did not allow a single exponential fit to
calculate the escape times for conformer A or conformer B at this temperature,
however, we found that the average folding time for folding trajectories
which did not go through the region of conformer B increased only modestly
to $t_f^{noB}=51,145$ MC sweeps from $t_f^{noB}=40384$ MC sweeps at the 
folding temperature
$T_f =230$K, while the folding time for trajectories going through 
the region of conformer B increased from $t_f^B=104,499$ MC sweeps to at least 
$t_f^B=173,856$ MC sweeps.
This demonstrates that with increasing glassiness of the system, it becomes
more and more difficult to escape the now much longer living traps.

To further understand the folding kinetics we studied the time evolution
of the fraction of unfolded states $P_{uf}(t)$ (as defined in Eq.~\ref{Puf})
versus time (in MC sweeps) for 
the four chosen temperatures. These fractions were calculated from the time
series of the 100 canonical Monte Carlo simulations of up to
200,000 sweeps for each of the four temperatures. Fig.~5 displays  
this quantity as a function of Monte Carlo time on a log-log scale. 
It is obvious that the time evolution of that quantity cannot be described
by a power law (which would imply a straight line in the plot) indicating
that the folding has to be described by a combination of exponentials or,
numerically simpler, a stretched exponential.
Indeed we found that for our two highest temperatures, the observed curves 
can be described by a  single exponential fit
\begin{equation}
P_{unfolded} = A \times e^{-t/\tau_1}~,
\end{equation}
while for the folding temperature
$T_f=230 $K and our lowest temperature $T=150$ K, which is below the
glass temperature, we  needed  stretched exponential fit:
\begin{equation}
P_{unfolded} = A \times e^{-(t/b)^c}
\end{equation}
to describe our data.  The lines in Fig.~5 
mark the fits through our data and we see that the chosen functional forms
describe well our data. Table~3 lists the coefficients of the chosen fits.
It is obvious from these  kinetic data that above the folding temperature 
no long living traps exist. This is in agreement with the smoothness of the
energy landscape which we observe in Fig.~1 and 2. As the temperature is 
lowered, around at temperature $T_f \approx 230$ K, non-exponential 
behavior started to be observed. In the beginning this mechanism can 
be described by a few exponentials indicating that  only small number 
of traps start to play a role. For instance, at the folding temperature
$T_f=230$ K, the kinetics could also be fitted well with a two exponential 
form (fit not shown).
This is consistent with Fig.~3 where we observe indeed only few local
traps in the free energy landscape.
However, as  the temperature gets lower and lower, the number of 
traps substantially increases and glassy-like dynamics is observed. 
 Fig.~4 shows that the many local minima separated by free energy 
barriers of many $k_BT$. As a result, the residence time in some local traps 
becomes of the order of the folding event. Folding dynamics is now
non-exponential (since different traps have different escape
times \cite{REF1}). In this regime stretched exponentials are much more
appropriate.  Such a behavior was predicted from
studies of minimal protein models and is now verified by us for a
realistic protein model.

To summarize, we have studied the thermodynamics and kinetics of the
peptide Met-enkephalin, using a combination of generalized-ensemble 
techniques and canonical Monte Carlo. Generalized-ensemble techniques
introduce an artificial dynamics and therefore do not allow to study 
directly the the kinetics of proteins. However, these sophisticated 
techniques enable reconstruction of the free-energy landscape of a protein.
We have shown that a careful analyses of these landscapes leads 
to indirect information on the folding kinetics. For this purpose,
we  compared our
generalized-ensemble results with dynamical Monte Carlo simulations at
appropriate temperatures. This  demonstrates the usefulness of
generalized-ensemble techniques in investigations of the mechanism
and kinetics of folding. Combining generalized-ensemble results with 
dynamical Monte Carlo simulations, the present study 
provide evidence that the concepts of folding that were developed 
in the past for minimalist models hold also for our peptide 
when simulated with an all-atom force field.

\newpage

\noindent
{\bf Acknowledgements}: \\
This work was supported by the National Science Foundation  under 
grants   \#CHE-9981874 and MCB-0084797. Part of this article was written 
 while U.H. was visitor at the Department of Physics at
the Bielefeld University. He thanks F.~Karsch for his hospitality
during his stay in Bielefeld.



\vfil

\newpage
{\huge Tables:}\\

\begin{table}[h]
\caption{Number of times that the ground state configuration was found
         in less than 200,000 MC sweeps and the lower limit for the average
         folding time (in MC sweeps). We further distinguish between the
         case that the ground state was found without first 
         visiting conformer `B' and the case where the ground state
         was found only after visiting conformer `B'}
\begin{center}
\begin{tabular}{||r|r|r|r|r|r|r||}
$T$[K] & $n_{f}$& $t_{f}$[MC sweeps]& $n_f^{no B}$
       & $t_f^{no B}$[MC sweeps] & $n_f^B$ & $t_f^B$ [MC sweeps]\\
\hline
1000 &  47 & 145599 (14286)& 40 & 80696 (12562) & 7 & 176145 (32753)\\
300  & 100 & 19864 (1141)& 45 & 8570 (759) & 55 & 29104 (2852)\\
230  &  86 & 77230 (7489)& 47 & 40384 (5505)& 39 & 104499(14479)\\
150  &  19 & 172866 (16964) & 14& 51145 (13132) & 5& 173856(34871)\\
\end{tabular}
\end{center}
\end{table}

\begin{table}[h]
\caption{Escape times for conformer A and B as function of temperature.
         For $T=150$ K no single exponential fit was possible.}
\begin{center}
\begin{tabular} {||r|r|r||}
$T$[K] & $\tau_{es}^A$ [MC sweeps]& $\tau_{es}^B$ [MC sweeps]\\
\hline
1000 & 9.0(5) & 6.9(6)\\
300  & 2022 (17) & 789 (16) \\
230  & 19430 (230) & 4830 (50)\\
150  & -           & - \\
\end{tabular}
\end{center}
\end{table}

\begin{table}[t]
\caption{Coefficients for a stretched  exponential fit 
         $ P_{unfolded} = A \times e^{-(t/b)^c}$
         of the frequency of unfolded
         conformations as a function of time (in MC sweeps) for 100
         canonical Monte Carlo runs. For the case of $T=1000$ K and $T=300$
         K we  added for comparison the coefficients for a 
         single-exponential fit
         (i.e. $c=1$) and mark these coefficients by a $*$.}
\begin{center}
\begin{tabular}{||r|r|r|r||}
$T$[K] & a & b & c   \\
\hline
  1000     & 1.002(3) &   310,000(3,000)   &  0.95(2)  \\
  1000     & 1(*)      & 304,250 (760) (*)  &  1 (*)\\
   300     & 1.004(8)  &    18,000(3,000)   &  0.91(15)\\
   300     & 1 (*)     & 20,200 (100) (*)   &  1 (*) \\
   230     & 0.97(2)  &    90000(4,000)    &  0.76(4)  \\
   150     & 1.005(1) &  6,613,000(20,000) &  0.43(1)  \\ 
 \end{tabular}
\end{center}
\end{table}
\ 
\\ \\ \\ \\ \\ \\ \\ \\ \\ \\ \\

\newpage
{\Large Figure Captions:}\\
\begin{enumerate}
\item{Fig.~1:} Free energy $G(O_A,O_B)$ as a function of both
                overlaps $O_A$ and $O_B$ (as defined in the text)
                for $T = 1000$ K. The data rely on a generalized-ensemble
               simulation of 1,000,000 sweeps. The contour lines are spaces
               $ 1k_B T$.
                Superimposed is the folding trajectory
               of a canonical simulation of 200,000 sweeps at the
               same temperature. We marked with ``U'' the region of 
               random unfolded conformers.
\item{Fig.~2:} Free energy $G(O_A,O_B)$ as a function of both
                overlaps $O_A$ and $O_B$ (as defined in the text)
                for $T = T_{\theta} =300$ K.  
               The data rely on a generalized-ensemble
               simulation of 1,000,000 sweeps.
               The contour lines are spaces $1 k_B T$.
                Superimposed is the folding trajectory
               of a canonical simulation of 200,000 sweeps at the
               same temperature. ``U'' marks random, unfolded conformers
               and ``A'' marks conformer A.
\item{Fig.~3:} Free energy $G(O_A,O_B)$ as a function of both
                overlaps $O_A$ and $O_B$ (as defined in the text)
                for $T = T_f =230$ K.  
                The data rely on a generalized-ensemble
               simulation of 1,000,000 sweeps.
               The contour lines are spaces $1 k_B T$.
                Superimposed is the folding trajectory
               of a canonical simulation of 200,000 sweeps at the
               same temperature. ``U'' marks random, unfolded structures,
               ``A'' conformer A and ``B'' marks conformer B.
\item{Fig.~4:} Free energy $G(O_A,O_B)$ as a function of both
                overlaps $O_A$ and $O_B$ (as defined in the text)
                for $T = 150$ K, well below the glass transition temperature
               $T_g =190 \pm 20$.  
               The data rely on a generalized-ensemble
               simulation of 1,000,000 sweeps.
               The contour lines are spaces $1 k_B T$.
                Superimposed is the folding trajectory
               of a canonical simulation of 200,000 sweeps at the
               same temperature. The minimum corresponding to the
               ground state (conformer A) is marked by ``A''.
\item{Fig.~5:} Probability of unfolded configurations as a function of
               Monte Carlo time on a log-log plot. The lines mark our
               fits (see text) through the data points.
\end{enumerate}

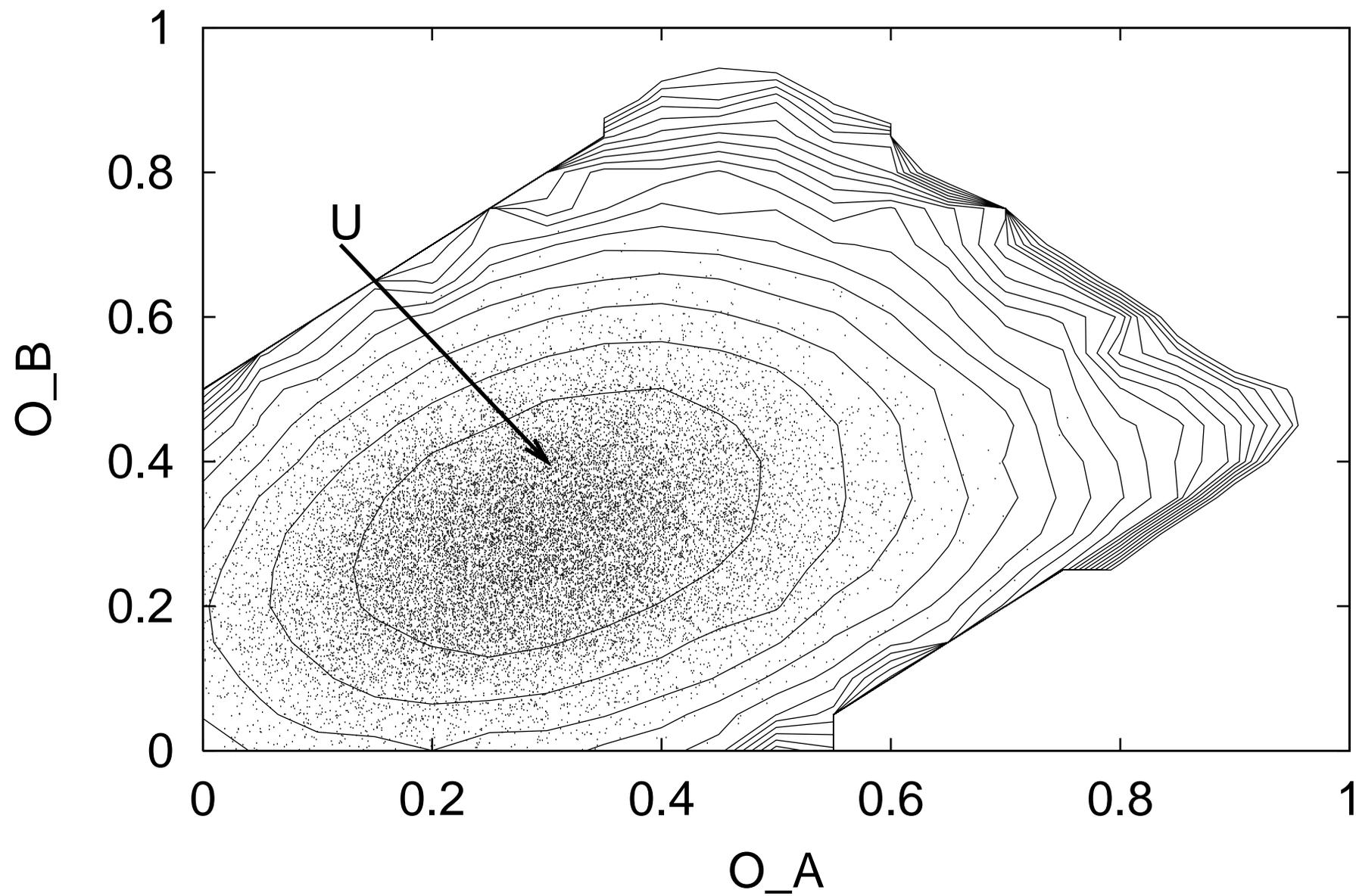
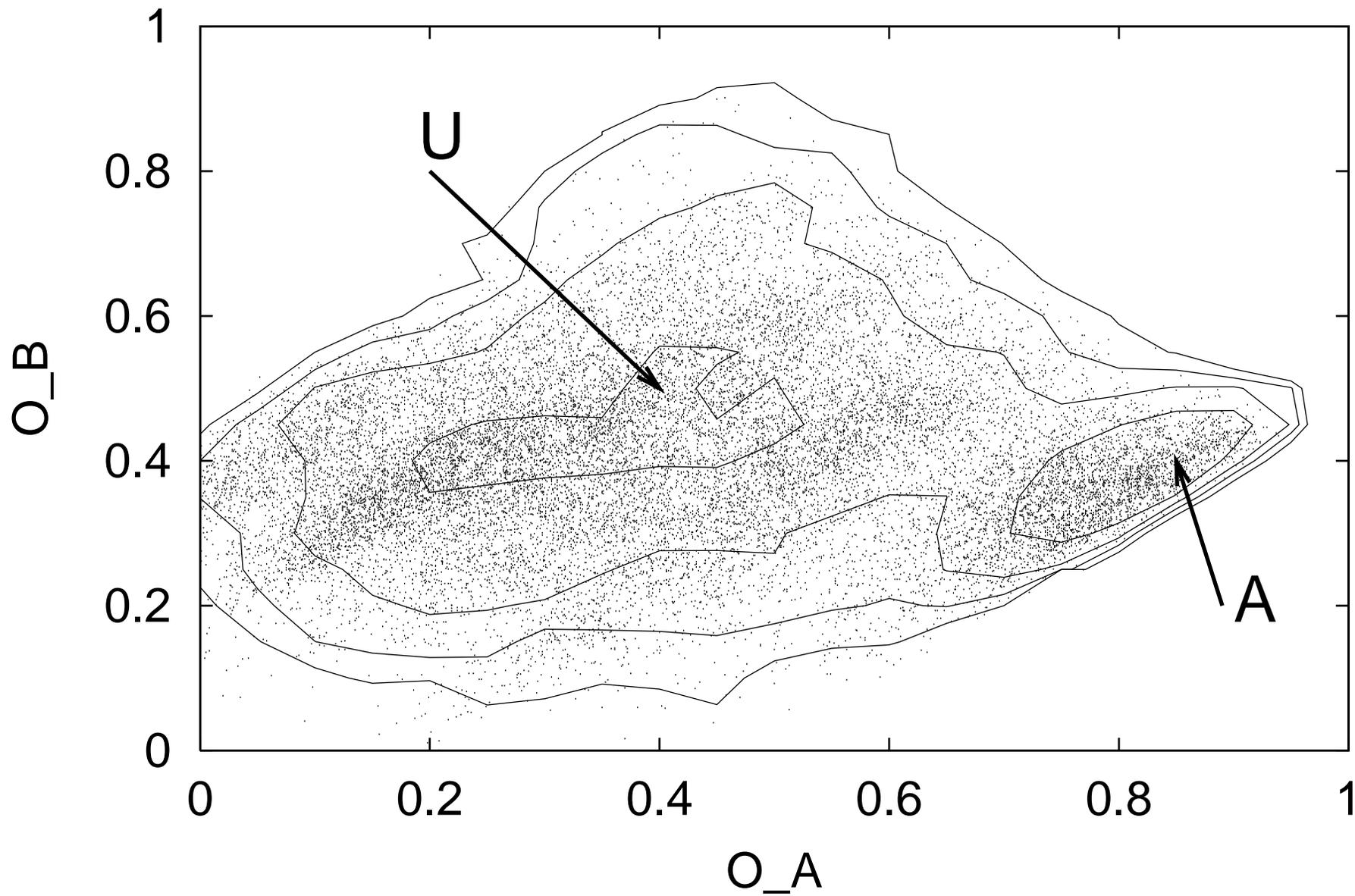
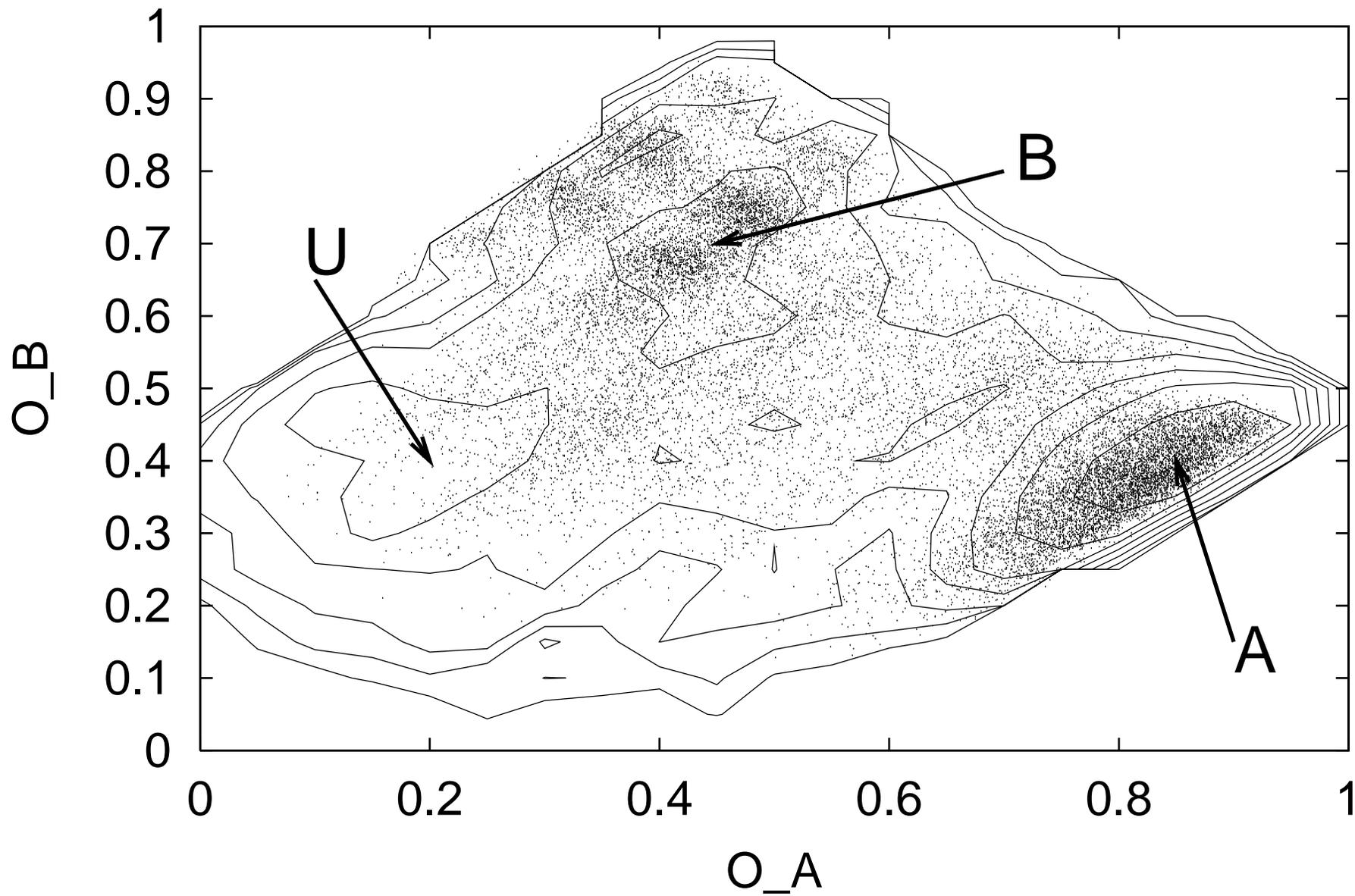
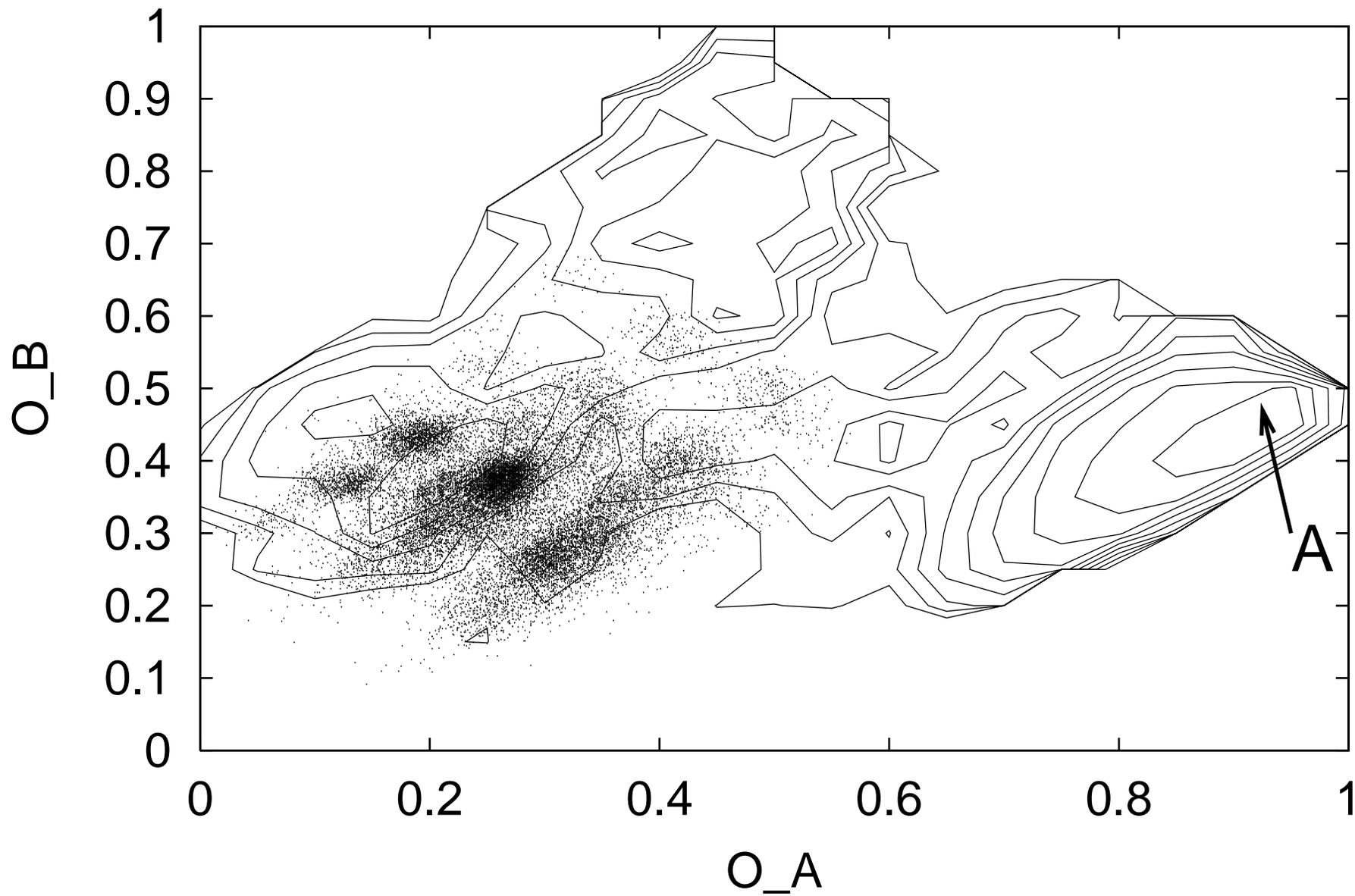
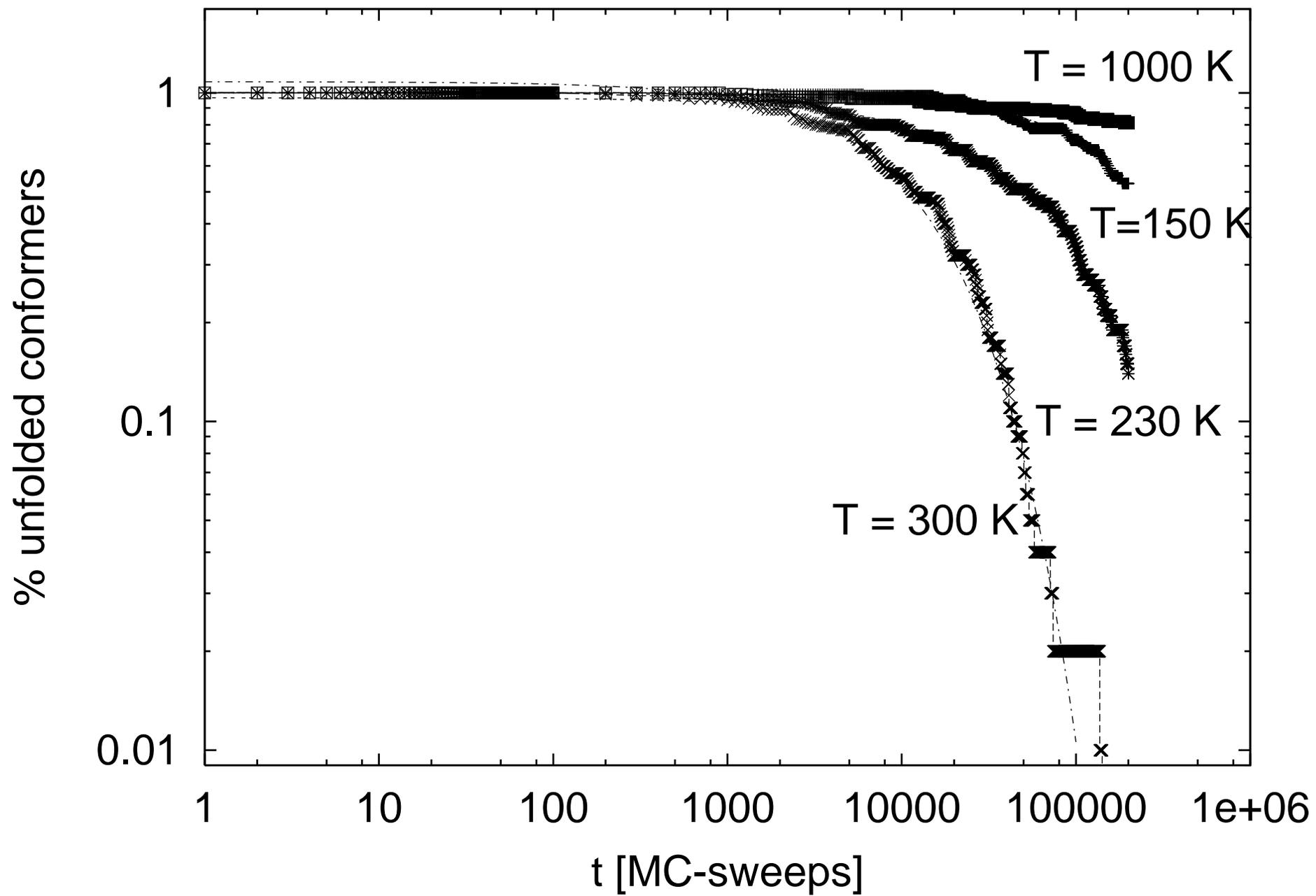
\end{document}